# Electrical transport in small bundles of single-walled carbon nanotubes: intertube interaction and effects of tube deformation

Taekyung Kim,[1] Gunn Kim,[2] Woon Ih Choi[3], Young-Kyun Kwon[2], and Jian-Min Zuo[4]

[1] The Center for Integrated Nanotechnologies (CINT), Sandia National Laboratories, Albuquerque, NM 87185
[2] Department of Physics and Research Institute for Basic Sciences, Kyung Hee University, Seoul, 130-701, Korea
[3] Computational Materials Science Group, National Renewable Energy Laboratory, Golden, Co 80401-3393
[4] Department of Materials Science and Engineering and the Frederick Seitz Materials Research Laboratory, University of Illinois at Urbana-Champaign, IL, 61801

**Abstract**

We report a combined electronic transport and structural characterization study of small carbon nanotube bundles in field-effect transistors (FET). The atomic structures of the bundles are determined by electron diffraction using an observation window built in the FET. The single-walled nanotube bundles exhibit electrical transport characteristics sensitively dependent on the structure of individual tubes, their arrangements in the bundle, deformation due to intertube interaction, and the orientation with respect to the gate electric field. Our *ab-initio* simulation shows that tube deformation in the bundle induces a bandgap opening in a metallic tube. These results show the importance of intertube interaction in electrical transport of bundled nanotubes.

PACS numbers: 85.35.Kt, 68.37.Lp, 73.22.-f

Single-wall carbon nanotube (SWNT) bundles formed by the van der Waals (vdW) interactions, are common in samples prepared by laser ablation.[1] Small bundles consisting of two or several parallel tubes are also common in chemical vapor deposition (CVD) growth.[2] A bundle can consist of metallic and/or semiconducting tubes. In fact, early studies of SWNTs were often performed using carbon nanotube (CNT) bundles.[3,4] Electronic and electrical transport in CNT bundles presents interesting features, such as pseudogaps[5,6,7,8] and gap openings[9] in armchair SWNTs, single electron transport[10] and metallic resistivity.[11] Although understanding these phenomena requires detailed atomistic structure about individual tubes, structural characterization together with the transport measurements has been accomplished only in a few cases for individual CNTs.[12,13] Here, we report a combined study of electrical transport and structural characterization of bundles comprising two tubes. Our results reveal that electronic transport in bundles can be dramatically influenced by proximity of neighboring tubes as well as by properties of individual nanotubes. According to our first-principles calculations, deformation of the tubes changes the band gap, which supports the experiment. Our computational results also demonstrate that the screening of the electric field by the tube bundle structure breaks symmetry of a SWNT, which enhances intertube interaction, leading to electrical transport characteristics that are different from the properties of individual nanotubes.

The structural characterization is made possible by a novel field-effect transistor (FET) device architecture with an observation window for transmission electron microscopy (TEM) shown in Fig. 1.[14] A thin slit is etched through a thin layer of Si as a TEM observation window. The Si layer also acts as the gate. SWNTs were grown across the slit on top of 250 nm thick of thermally grown $SiO_2$ by CVD using ultrathin metal film catalysts.[2,15] The metal catalyst layer used here is Mo/Fe/Al of 0.5/1.0/8.0 nm thickness. Growth temperature was 900 $^{\circ}$C and the gas mixture was $CH_4$ and $H_2$. The details of growth process are described elsewhere.[2] After CNT growth, electrodes were formed by conventional optical lithography using Au/Ti.

Figure 2 shows two examples of current-voltage ($I_D$-$V_G$) curves for FET devices with two-tube bundle structures. In Fig. 2(a), the $I_D$-$V_G$ behavior is similar to that of a device made of two separated, parallel tubes. The two tubes in the device (Device I) shown in the inset of Fig. 2(a) were separated on the source contact. The $I_D$-$V_G$ curve in this case is a characteristics of a mixture of semiconducting and metallic SWNTs. Two tubes (Device II) are bundled at both the source and drain contacts. This device is neither fully on nor fully off and its conductance is only slightly modulated by the gate bias.

To understand the $I_D$-$V_G$ curves shown in Fig. 2, we carried out structure characterization using a JEOL 2010F transmission electron microscope (TEM) operated at a high voltage of 200 kV. The details of structure characterization were published in Refs. 16 and 17. Briefly, diffraction patterns (DPs) were recorded from the bundles using the nano-area electron diffraction (NED) technique and a parallel electron beam of 50 nm in diameter. All diffraction experiments were carried out after electrical transport measurements to avoid any damage by high energy electrons to CNTs.

An analysis of the DPs was carried out as follows. The positions of the layer lines (indicated as (0,1) or (1,0) in Fig. 3) were used to measure tube chiral angles. A measurement error was ~0.1°. The equatorial line intensity was used to measure the tube diameters and the distance between two tubes in projection along the incident electron beam. The two diameters ($D_1$ and $D_2$) and the separation distance ($d$) are used as parameters to fit the equatorial line oscillations using $\left|D_1 J_o\left(\pi R D_1\right) + D_2 J_o\left(\pi R D_2\right)\exp(2\pi i R d)\right|^2$,[17] where $J_0$ is the zeroth order Bessel function, $R$ is the reciprocal lattice vector, and $d$ is the peak distance. On the basis of the tube diameters and the chiral angles found in the analysis, we chose the chiral indices. Independently, we also performed the layer line fitting with high order Bessel functions[16] to confirm the choice of chiral indices from the aforementioned analysis. For Fig. 3(a) recorded from the device of Fig. 2(b), two chiral indices were determined to be (17,5) and (19,11). The separation distance was $d$ = 0.05 nm in Fig. 3(b).[17] The intensity oscillation is extremely sensitive to $d$, which can be measured within an accuracy of 0.01 nm. Note that there is a difference between experimentally measured diameter ($D_E$) and theoretical one ($D_T$).

For the (17,5) CNT, $D_E$ and $D_T$ are 1.60 nm and 1.56 nm, respectively, and for the (19,11) CNT, 2.35 nm, and 2.05 nm, respectively. The experimental diameters are slightly larger than the theoretical values and the difference is more pronounced for (19,11) tube, which is the larger of two tubes. Interestingly, the two SWNTs in Device II are almost vertically stacked with respect to the gate oxide. Device I in Fig. 2(a) consists of a semiconducting (23,10) CNT and a metallic (25,16) CNT.

Among the two devices, the $I_D$-$V_G$ characteristic of Fig. 2(a) has a nonzero minimum and a large current at negative bias, which is consistent with a device made of a semiconducting and metallic tubes, which is in agreement with the structural determination by diffraction. The charges are injected and carried into the two tubes independently. When two tubes are not structurally overlapped with respect to the electric field of the gate, the two tubes act as independent channels. In this case, the conductance of the semiconducting channel is independently saturated (on-state) and minimized (off-state). In contrast, the current level of Device II in Fig. 2(b) is neither in the "on"-state (saturated) nor in the "off"-state (minimized) and its conductance is slightly modulated by the gate bias. This device consists of metallic and semiconducting tubes bundled at both source and drain contacts. The components of active channels of Device I and II are the same; a metallic and a semiconducting SWNTs. However, the structural configuration at the contacts with respect to the electric field by the gate voltage is different. The details are described below.

The observed device characteristics cannot be explained with just the properties of individual tubes. If each tube were to conduct separately, the tube with largest conductance would dominate.[18] In Fig. 2(a), the metallic tube is the only conducting channel, when the semiconducting tube is in off-state. ~75 % of the total current is transported through the metallic tube, showing that the metallic tube is the dominant channel, albeit with a significant contribution (~25 %, ~200 nA of current modulation out of ~820 nA of total current) from the semiconducting tube. For Device II, the channel conductance is slightly modulated by the gate bias, indicating that the active channel is neither pure metallic nor semiconducting. Such variation in the metallic behavior between the various bundles is certainly unexpected. The metal contacts (Au/Ti) were formed by deposition without heat treatment. This avoids the possibility of metal carbide formation.[19,20]

If two well-contacted tubes acted as two independent channels, the $I_D$-$V_G$ characteristic of Device II would be similar to that of Device I. To understand the transport properties of small tube bundles in a FET, we consider several effects due to point defects in CNTs, the contact with electrodes, the tube deformation by vdW forces, and the electric field by the gate bias. Point defects on a metallic SWNT may break the mirror symmetry of the tube and alter the electronic structure.[21] While the electron diffraction is consistent with the average CNT structure, we can not rule out a possibility of the presence of randomly distributed point defects in the metallic tube. We note that the current modulation of Device II (~25% in the measured gate voltage range) is comparable to that in the Ref 21 (~30% in the measured range). The electronic structure in a CNT can be also modified by its contact with metals or gate oxide.[22,23,24,25] X.-F. Li *et al*. investigated the transport properties of a device with two SWNTs attached to metal electrodes and found that a device consisting of two semicondcuting (8,0) SWNT can show metallic transport characteristic, depending on the intertube distance and metal contact coupling.[26] In a CNT bundle, CNTs can be deformed along the radial direction by vdW forces.[27] This radial deformation has been confirmed and reported by ED analysis.[17] The radial deformation breaks the cylindrical symmetry and changes the electronic structure in the CNT. Finally, we consider the effect of electric field by the gate bias. When CNTs are stacked, the electric field can be partly screened by adjacent nanotubes . Therefore, the symmetry of the two-tube bundle is broken and the tube bundle experiences a bandgap change. For Device II, the screening effect is expected to be strong whereas for Device I, no screening effect is expected.

To see the effect of tube deformation and electric field shielding, we performed *ab initio* electronic structure calculations for two-tube bundles with radial deformation[17] using the density functional theory (DFT) within the local density approximation (LDA). The electron-ion interaction was described by norm-conserving Kleinman-Bylander pseudopotentials[28,29] and the wave functions were expanded using a numerical atomic orbital basis set (double zeta and polarization) with a mesh cutoff of 230 Ry in the SIESTA code.[30,31] Since DFT does not describe the vdW interaction well, we constructed an initial structure two-tube bundle with ~5 percent deformation as confirmed by the diffraction analysis,[17] and minimized the total energy and atomic forces of the model structure. The force minimization by the conjugate gradient relaxation was done until the atomic forces are smaller than 0.02 eV/Å. Because chiral tubes have an unacceptably large unit cell, we chose achiral zigzag CNTs with similar diameters. CNTs with different chiralities but almost the same diameter are expected to show similar electronic features with respect to the band gap. In our calculations, (26,0)-(21,0) SWNTs containing 188 carbon atoms in the super cell were used, and the intertube distance between nanotubes in the same bundle changed from the initial value of 3.35 Å (the interlayer distance of graphite) to around 3.0 Å. From the calculation, we found a significant change in the band gap with tube deformation. The band gap of the deformed CNT bundle was 18 meV as shown in Fig. 4(a). The band gaps of the (26,0) and (21,0) CNTs with no deformation were 372 and 7 meV, respectively. Thus, the theory predicts a larger band gap for the (21,0) CNT from to radial deformation induced by the vdW intertube interaction. According to our calculations, the external electric field gives rise to symmetry breaking due to the electrostatic screening in the region between the two neighboring nanotubes. We note that the weak, but observable, gate dependence of FET suggests somewhat a larger gap than the theory predicts. In the calculation, however, the interaction between metal contacts and CNTs was not considered.

In conclusion, our results of electronic transport and structure characterization show that nanotube bundles exhibit unusual electrical transport characteristics due to the combination of several effects from the radial deformation induced by vdW forces, and screening of the electric field by the gate bias, depending on tube configurations, and interaction between CNTs and metal contacts. Theoretical calculations indicate that the change in electrical behavior originates from a change in the electronic structure of the tube, which is due to proximity of neighboring tubes and their radial deformation.

This work was supported by U.S. Department of Energy Grant DEFG02-01ER45923 and DEFG02-03ER46095. TK also acknowledges the supported by Laboratory Directed Research and Development, Sandia is a multiprogram laboratory operated by Sandia Corporation, a Lockheed Martin Company, for the United States Department of Energy under contract DE-AC04-94AL85000. This work was also supported by U.S. Department of Energy Grant DEFG02-01ER45923 and DEFG02-03ER46095. GK appreciates the financial support by a grant from Kyung Hee University in 2009. YK was supported by NRF of Korea grant KRF-2009-0074951. Microscopy was carried out at the Center for Microanalysis of Materials at the Frederick Seitz Materials Research Laboratory, which is partially supported by the U.S. Department of Energy under grant DEFG02-91-ER45439. We also thank Prof. Rogers, Prof. Shim, and Prof. Khang for help with CNT growth and helpful discussion and Dr. Olson and Prof. Petrov for TEM holder.

Figure Captions

Figure 1 Scanning electron microscopy (SEM) image of a SWNT bundle in the FET, enlarged view from the circled area in the inset. Inset shows a large view of the device array fabricated on a Silicon-On-Insulator (SOI) wafer.

Figure 2 Current-Voltage ($I_D$-$V_G$) characteristics of two-tube bundles. (a) Current modulation is superimposed onto the metallic contribution from Device I. The inset shows two tubes are separated at the source contct. (b) $I_D$-$V_G$ characteristics obtained from Device II. The dotted arrow in the inset represents the direction of the electric field during device operation. The two SWNTs are almost vertically stacked with respect to the electric field. All the $I_D$-$V_G$ measurements were performed at 0.5 V of source-drain voltage.

Figure 3 (a) Diffraction pattern from experiment (left) and simulation (right). Both diffraction spots and equatorial oscillation match well. (b) Comparison of intensity profiles of the equatorial lines obtained from experiment and simulation. The best match was found with $d = 0.05$ nm.

Figure 4 (a) Band structure of a small bundle consisting of (26,0)-(21,0) CNTs. The inset shows the band gap opening of the (21,0) CNT. The band gap of (21,0) becomes 18 meV, The original band gap of this CNT is 7 meV. Because of the vdW interaction, the carbon nanotubes are deformed. (b) A model structure used in the calculation is shown. The model includes ~5% radial deformation for the calculation.

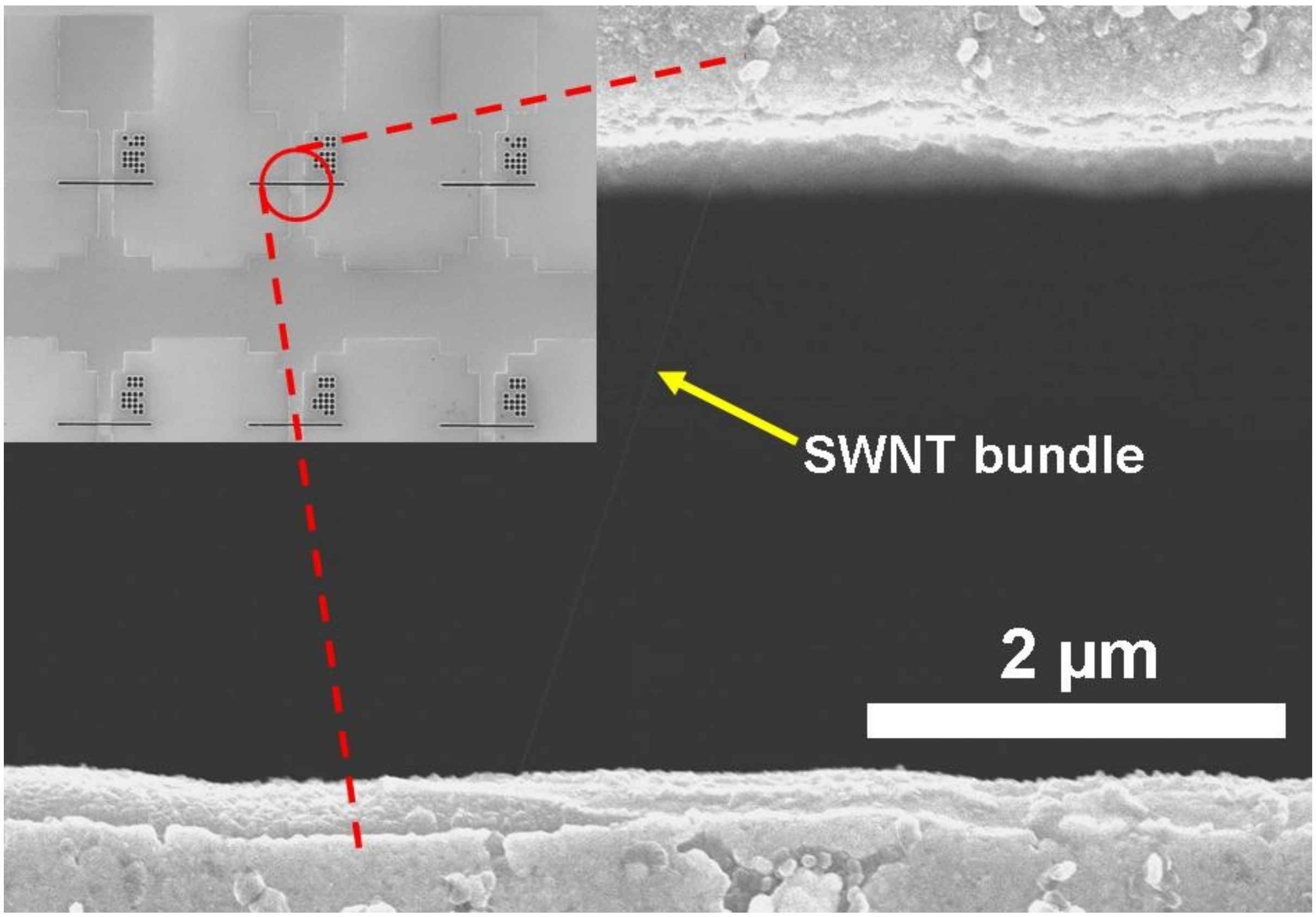

Figure 1

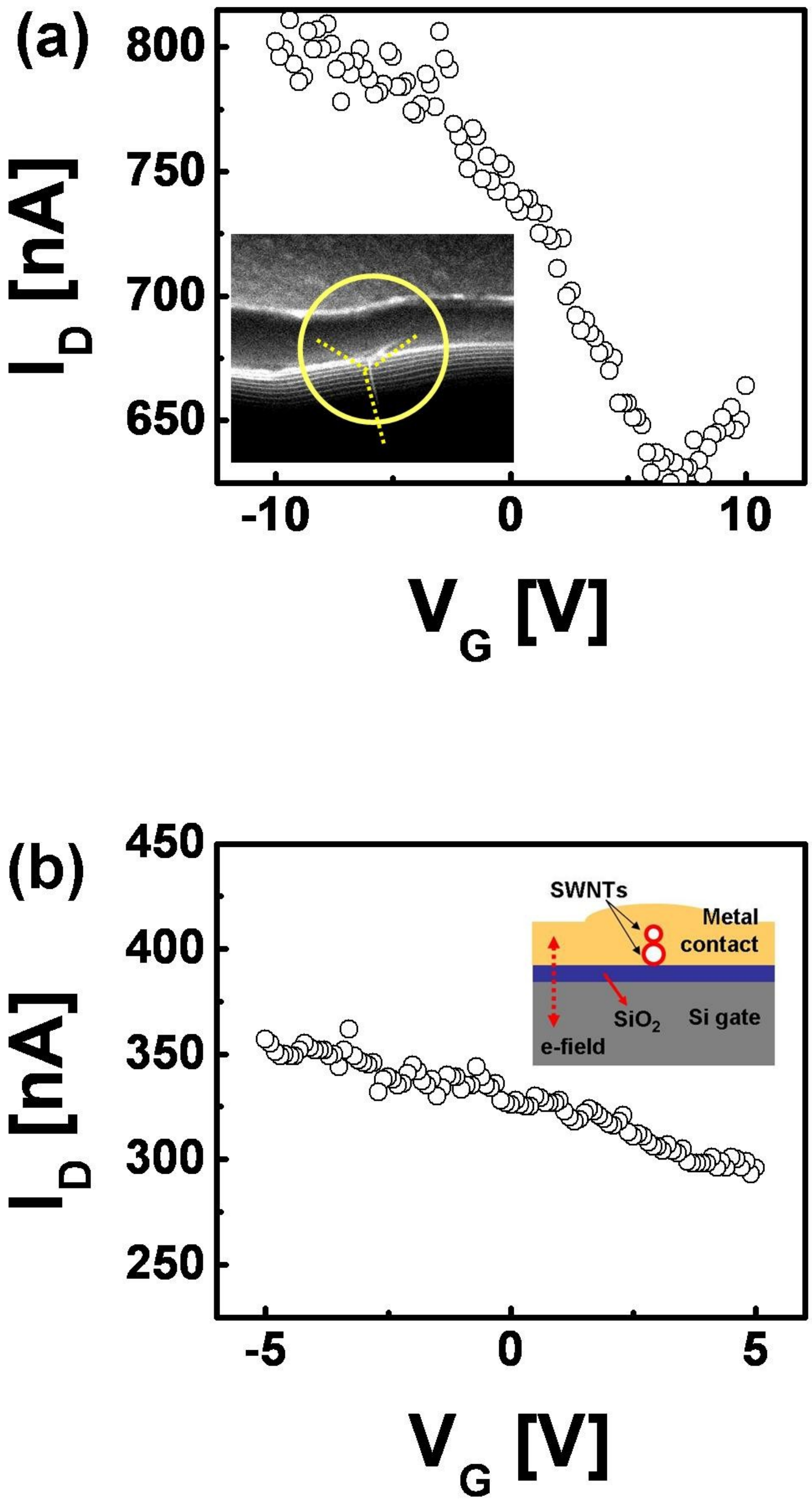

(a)
800
750
700
650
-10
0
10
$I_D$ [nA]
$V_G$ [V]
(b)
450
400
350
300
250
-5
0
5
SWNTs
Metal
contact
$SiO_2$
Si gate
e-field
$I_D$ [nA]
$V_G$ [V]

Figure 2

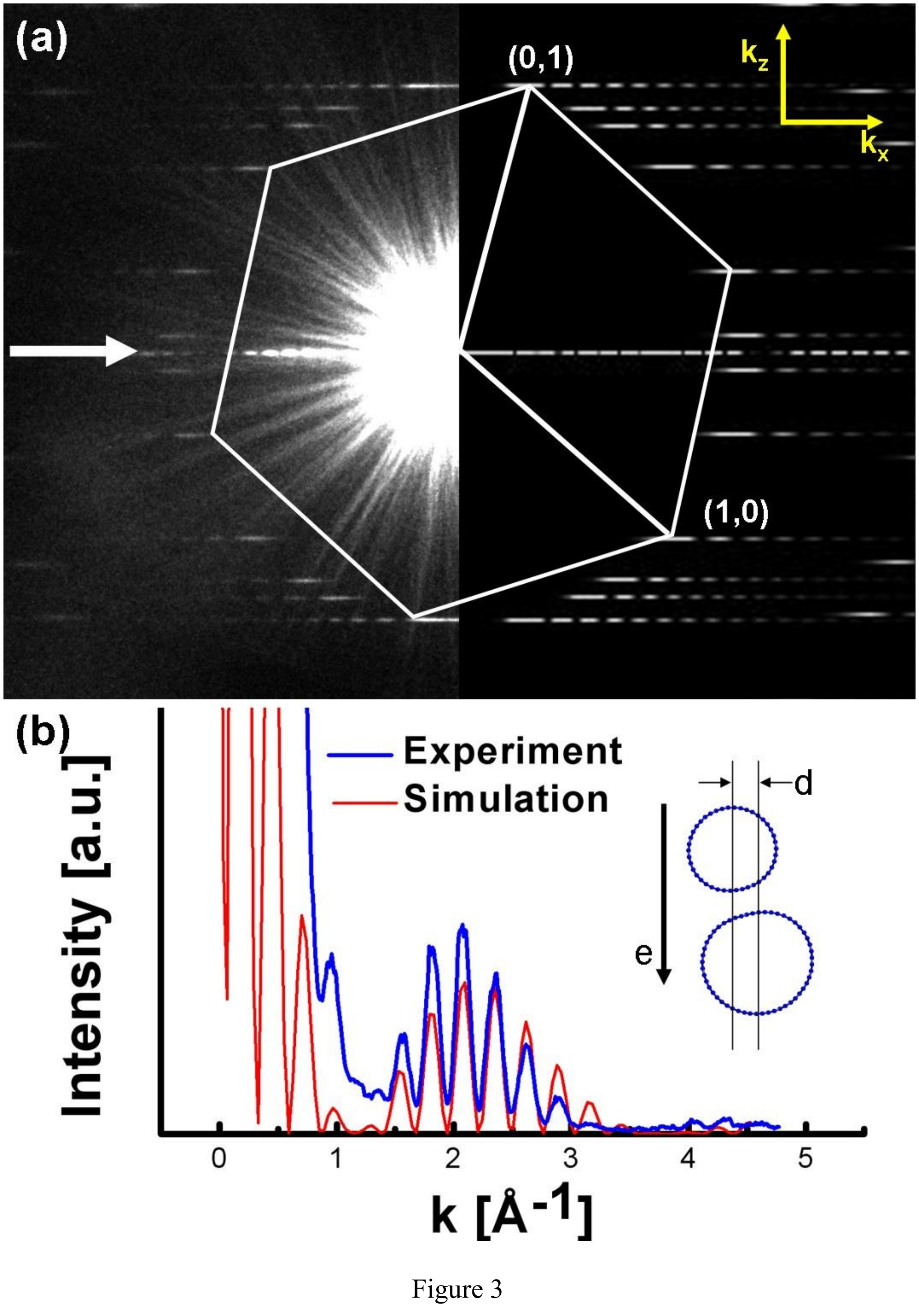
(a)
(0,1)
(1,0)
$k_z$
$k_x$
(b)
Intensity [a.u.]
Experiment
Simulation
d
e
0
1
2
3
4
5
k [Å$^{-1}$]

Figure 3

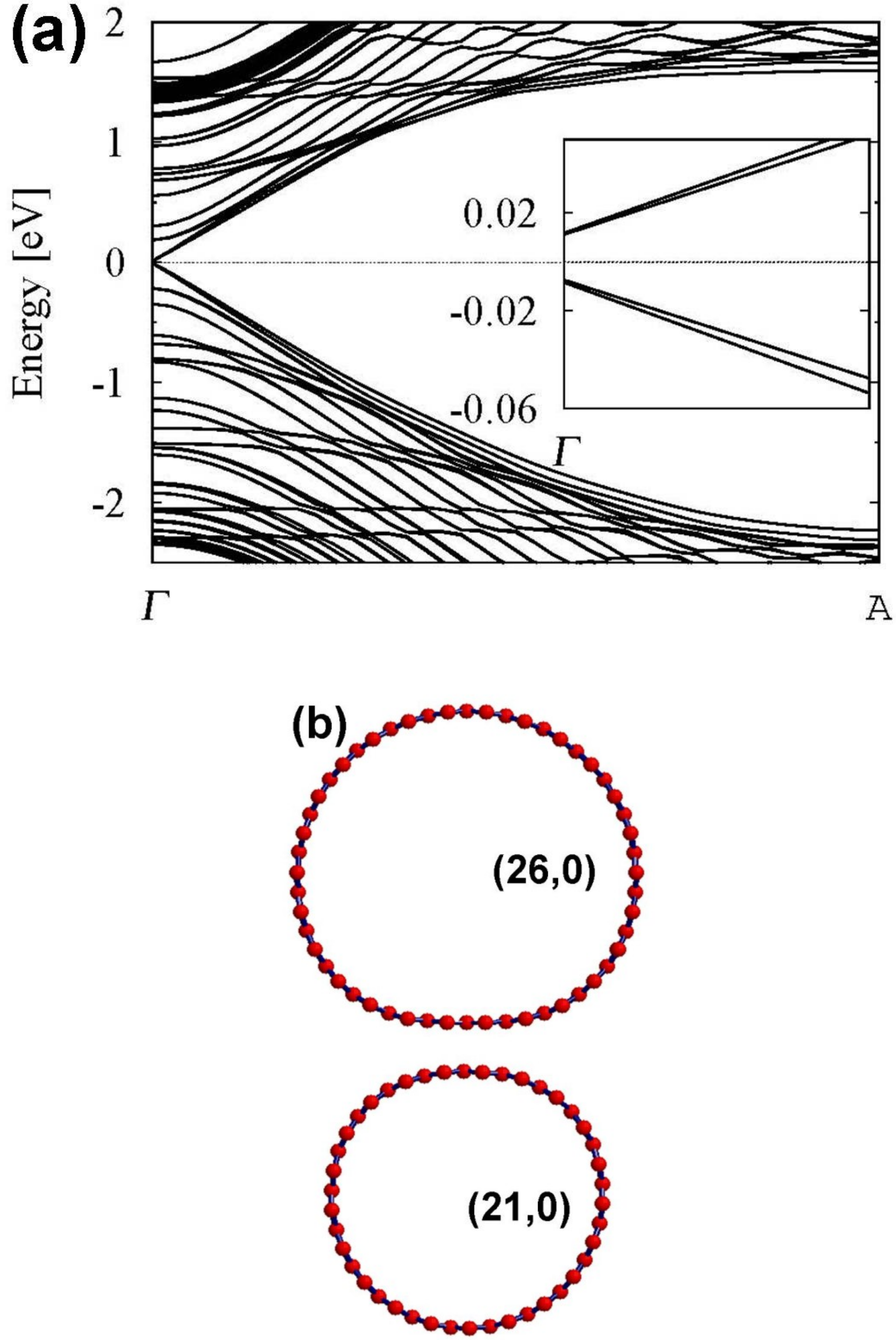

Figure 4